\documentclass[amsmath,amssymb,floatfix,reprint,twocolumn,superscriptaddress,longbibliography,prl]{revtex4-1}

\usepackage{graphicx}
\usepackage{epsfig}
\usepackage{hyperref}
\usepackage{xcolor}
\usepackage{bm}

\begin{document}

\title{Radio-frequency manipulation of state populations in an entangled fluorine-muon-fluorine system}

\author{David~Billington}
\email{billingtond1@cardiff.ac.uk}
\affiliation{School of Physics and Astronomy, Cardiff University, Queen's Building, The Parade, Cardiff, CF24 3AA, United Kingdom}
\author{Edward~Riordan}
\affiliation{School of Physics and Astronomy, Cardiff University, Queen's Building, The Parade, Cardiff, CF24 3AA, United Kingdom}
\author{Majdi~Salman}
\affiliation{School of Physics and Astronomy, Cardiff University, Queen's Building, The Parade, Cardiff, CF24 3AA, United Kingdom}
\author{Daniel~Margineda}
\affiliation{School of Physics and Astronomy, Cardiff University, Queen's Building, The Parade, Cardiff, CF24 3AA, United Kingdom}
\author{George~J.W.~Gill}
\affiliation{School of Physics and Astronomy, Cardiff University, Queen's Building, The Parade, Cardiff, CF24 3AA, United Kingdom}
\author{Stephen~P.~Cottrell}
\affiliation{ISIS Facility, Rutherford Appleton Laboratory, Harwell Campus, Didcot, Oxon, OX11 0QX, United Kingdom}
\author{Iain~McKenzie}
\affiliation{TRIUMF, Vancouver, V6T 2A3, Canada}
\author{Tom~Lancaster}
\affiliation{Department of Physics, Center for Materials Physics, Durham University, Durham DH1 3LE, United Kingdom}
\author{Michael~J.~Graf}
\affiliation{Department of Physics, Boston College, Chestnut Hill, Massachusetts 02467, USA}
\author{Sean~R.~Giblin}
\email{giblinsr@cardiff.ac.uk}
\affiliation{School of Physics and Astronomy, Cardiff University, Queen's Building, The Parade, Cardiff, CF24 3AA, United Kingdom}

\date{\today}

\begin{abstract}
Entangled spin states are created by implanting muons into single crystal LiY$_{0.95}$Ho$_{0.05}$F$_{4}$ to form a cluster of correlated, dipole-coupled local magnetic moments.
The resulting states have well-defined energy levels allowing experimental manipulation of the state populations by electromagnetic excitation.
Experimental control of the evolution of the muon spin polarization is demonstrated through application of continuous, radio-frequency magnetic excitation fields.
A semiclassical model of quantum, dipole-coupled spins interacting with a classical, oscillating magnetic field accounts for the muon spin evolution.
On application of the excitation field, this model shows how changes in the state populations lead to the experimentally observed effects,
thus enabling a spectroscopic probe of entangled spin states with muons.
\end{abstract}

\maketitle

The preparation, manipulation and measurement of entangled quantum states lies at the heart of emerging quantum technologies.
There are numerous methods for creating entangled states, where the manipulation is enabled by electromagnetic pumping at a frequency corresponding to the interval between the energy levels.
One such entangled state often occurs when a positively charged, $100\%$ spin-polarized muon, $\mu^{+}$, is implanted into a material containing fluorine.
The stopped $\mu^{+}$ tends to form hydrogen-like bonds with two nearby F$^{-}$ ions forming a linear ($180^{\circ}$ bond angle) F--$\mu$--F complex.
Magnetic dipole coupling between the $\mu^{+}$ spin ($S=1/2$) and the two F$^{-}$ nuclear spins ($I=1/2$) produces four doubly-degenerate, so-called F--$\mu$--F energy eigenstates \cite{brewer:86,lancaster:07,lancaster:09}.
The resulting $\mu^{+}$ spin polarization, as measured by muon spin relaxation ($\mu^{+}$SR), evolves over time because the initial state of the system is, essentially, a statistical mixture of superpositions of (typically) non-degenerate energy eigenstates, and exhibits oscillations driven by the magnetic dipole coupling between the entangled moments comprising the F--$\mu$--F states \cite{muonbook}.

In recent years, there has been significant effort to increase the decoherence time of entangled states as this is an obstacle to enhanced quantum technologies.
In general, the environment of a quantum system acts as a source of decoherence \cite{zurek:03} where the initial quantum information describing the system leaks into the environment and can no longer be recovered.
Recently, the decoherence of the entangled F--$\mu$--F states was successfully modeled by considering interactions with the {\it local} (nearest-neighbor) magnetic moments of the F--$\mu$--F complex \cite{wilkinson:20,wilkinson:21}, demonstrating that the F--$\mu$--F states are robust and well isolated from other, {\it external} environmental dephasing mechanisms.
In this letter, we demonstrate experimental control of the evolution of the $\mu^{+}$ spin polarization in single crystal LiY$_{0.95}$Ho$_{0.05}$F$_{4}$ and demonstrate that the spin system is effectively isolated from the external environment over the lifetime of the measurement, thus enabling the experimental manipulation of entangled state populations.
Control of the transitions between the entangled spin states is achieved simply by {\it in situ} radio-frequency (RF) electromagnetic excitation within the $\mu^{+}$SR sample environment.
Previous RF-$\mu^{+}$SR experiments have focused on either decoupling the $\mu^{+}$ spin from its local environment \cite{cottrell:98}, or exploiting the isolated bipartite muonium system (which does couple to its local environment) in order to investigate the local coupling constants \cite{blazey:86}.
In contrast, here the RF excitation field is applied continuously to drive transitions between the entangled F--$\mu$--F states.
Although there are more precise ways of manipulating state populations through careful timing and shaping of RF pulses, the continuous application of the RF-field is shown to be sufficient for this proof-of-principle experiment.

\begin{figure*}[t!]
\centerline{\includegraphics[width=1.0\linewidth]{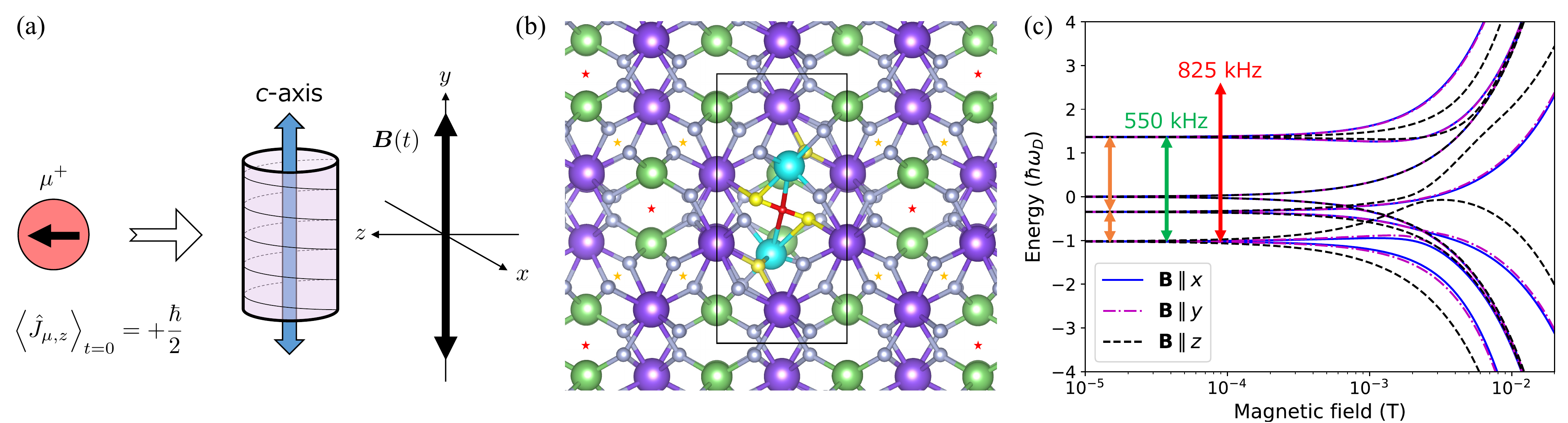}}
\caption{(a) Schematic of the experimental geometry.
(b) Crystal structure of LiYF$_{4}$ in the tetragonal $ac$-plane (experimental $zy$-plane) with the Li$^{+}$ (green), Y$^{3+}$ (purple) and F$^{-}$ (grey) ions shown as spheres.
The calculated $\mu^{+}$ stopping site (red) and associated structural distortion (described in the main text) is shown in the center.
Highlighted are the two nearest-neighbor F$^{-}$ ions (yellow), the two next-nearest-neighbor Li$^{+}$ ions (cyan) and the two next-next-nearest-neighbor F$^{-}$ ions (yellow) of $\mu^{+}$ which are included in the $\mu$F$_{2}$Li$_{2}$F$_{2}$ cluster calculation and whose positions are most affected by the structural distortion.
Other crystallographically-equivalent $\mu^{+}$ stopping sites are indicated by red and orange stars for sites with the same $\mu$F$_{2}$Li$_{2}$F$_{2}$ atomic geometry as the one shown and for sites with the same atomic geometry but rotated by $90^{\circ}$ about the $c$-axis, respectively.
The black rectangle is the tetragonal unit cell.
(c) Energy level diagram of the F--$\mu$--F complex shown in (b) as a function of magnetic field.
For small ${\bm B}$-fields, three transitions are possible (orange and green double-ended arrows).
The $550$~kHz transition (green) used in this experiment is highlighted, along with an $825$~kHz excitation (red) which is also used but, importantly, does not correspond to a transition.
For the $\mu^{+}$ stopping sites indicated by orange stars in (b), $x$ and $z$ are exchanged in (c).
}
\label{fig1}
\end{figure*}

The initial states are created by implanting $100\%$ spin polarized $\mu^{+}$ into single crystal LiY$_{0.95}$Ho$_{0.05}$F$_{4}$ where they predominantly stop near two F$^{-}$ ions and form entangled states similar to the F--$\mu$--F states.
We used a previously-studied large single crystal of LiY$_{0.95}$Ho$_{0.05}$F$_{4}$ \cite{johnson:11} which has a tetragonal structure that is essentially unchanged from the parent LiYF$_{4}$ and is known to have more than $80\%$ of the implanted $\mu^{+}$ form entangled states.
While the Ho$^{3+}$ ion is strongly magnetic, in LiY$_{1-x}$Ho$_{x}$F$_{4}$ it was shown that the dipole-coupled oscillations at $T=50$~K
were independent of Ho$^{3+}$ concentration for $0.002\leq x\leq0.086$ \cite{johnson:11},
demonstrating that the Ho$^{3+}$ magnetic moments are fluctuating very rapidly and thus on a timescale far outside the $\mu^{+}$ time window,
so will have a negligible effect on the measurements reported here.

To demonstrate experimentally how the $\mu^{+}$ polarization is sensitive to the population of the underlying eigenstates, and that we can manipulate and observe changes in these populations, $\mu^{+}$SR spectra were measured using the EMU spectrometer~\cite{giblin:14} at the STFC-ISIS facility, UK.
The $\mu^{+}$ spins are initially polarized along $+z$, and the RF coil was oriented such that the RF excitation field, ${\bm B}(t)=[0,B_{y}\cos(\omega_{\rm c}t),0]$, was linearly polarized along $y$ (perpendicular to the incoming $\mu^{+}$ beam, $z$) with the LiY$_{0.95}$Ho$_{0.05}$F$_{4}$ single crystal aligned with its $c$-axis along the field axis [see Fig.~\ref{fig1}(a)].
All of our measurements were performed at temperature $T=50$~K with zero static ${\bm B}$-field.
Only the oscillating ${\bm B}(t)$ was applied continuously over about $15$ $\mu^{+}$ lifetimes (about $33$~$\mu$s, the lifetime of the measurement) using a RF saddle coil in close proximity to the sample (details in Ref.~\cite{supp}).
A nominal excitation frequency [$\omega_{\rm on}/(2\pi)=550$~kHz] was chosen to match the highest-energy transition between the energy levels of the F--$\mu$--F states [see Fig.~\ref{fig1}(c)] which, hereafter, we refer to as the on-resonance condition.
Spectra were measured both on and off [$\omega_{\rm off}/(2\pi)=825$~kHz] the resonance condition and, for each RF frequency, data was collected alternately with (RF-on, $B_{y}\neq0$) and without (RF-off, $B_{y}=0$) the applied excitation field for $500$ accelerator pulses each, where each pulse duration is about $80$~ns (full-width-half-maximum).

To determine the $\mu^{+}$ stopping site and associated local structural distortion within the crystal \cite{moller:13,wilkinson:20,wilkinson:21} first-principles calculations \cite{castep,mufinder1,mufinder2} were performed.
We found a unique minimum energy structural configuration in which the two nearest-neighbors of the $\mu^{+}$ are both F$^{-}$ ions
(forming a state resembling F--$\mu$--F) the two next-nearest-neighbors are both Li$^{+}$ ions and the two next-next-nearest-neighbors are both F$^{-}$.
Together, these comprise a $\mu$F$_{2}$Li$_{2}$F$_{2}$ cluster, as shown in Fig.~\ref{fig1}(b).
There are $16$ symmetry-equivalent positions within the tetragonal structure, however, since rotations of the F--$\mu$--F bonds about the $z$-axis (initial $\mu^{+}$ spin direction) do not affect the observed $\mu^{+}$ polarization, there are only two orientations (rotated by $90^{\circ}$ about the $c$-axis) of the crystallographically equivalent $\mu^{+}$ stopping site to consider.
The calculations show that the F$^{-}$ ions and Li$^{+}$ ions move towards and away from the $\mu^{+}$ site, respectively, to minimize the electrostatic energy.
Table~S1 in Ref.~\cite{supp} lists the calculated atomic positions for the two orientations of the $\mu^{+}$ stopping site.

To model the decoherence of the F--$\mu$--F states in LiY$_{0.95}$Ho$_{0.05}$F$_{4}$, magnetic dipole interactions with the two next-nearest Li$^{+}$ and two next-next-nearest F$^{-}$ ions must be included.
From the calculated geometry, we construct and optimize \cite{storn:97} a dipole interaction Hamiltonian for a $\mu$F$_{2}$Li$_{2}$F$_{2}$ cluster to which we add a term representing the interaction of the spins with an oscillating magnetic field, ${\bm B}(t)$, with frequency equal to the level separation.
Utilizing the Quantum Toolbox in Python (QuTiP) module \cite{johansson:12,johansson:13}, we then evolve the initial density matrix according to our optimized Hamiltonian and evaluate the expectation value of the $z$-direction $\mu^{+}$ spin projection operator, $\big\langle\hat{J}_{\mu,z}\big\rangle(t)$.

The only stable isotope of fluorine is $^{19}$F [$I=1/2$, $\gamma_{\rm F}/(2\pi)=40.053$~MHz\,T$^{-1}$] and the stable isotopes of lithium are $^{6}$Li [$I=1$, $\gamma_{^{6}{\rm Li}}/(2\pi)=6.2661$~MHz\,T$^{-1}$] and $^{7}$Li [$I=3/2$, $\gamma_{^{7}{\rm Li}}/(2\pi)=16.548$~MHz\,T$^{-1}$]
which occur with relative abundances of $7.42\%$ and $92.58\%$, respectively.
Because $^{7}$Li is much more abundant and the gyromagnetic ratio of $^{6}$Li is about $1/3$ smaller, only $^{7}$Li is considered in our model.
The initial density matrix for the $\mu$F$_{2}$Li$_{2}$F$_{2}$ cluster, $\hat{\rho}_{\mu{\rm F_{2}Li_{2}F_{2}}}(t=0)$, is given by the tensor product of the initial density matrices of the seven subsystems.
Except for $\mu^{+}$, which is initially polarized in the $\left|\uparrow\right\rangle$ state, all of the other initial density
matrices are (in the $\hat{J}_{z}$-basis) identity matrices weighted with equal probability,
$\hat{\rho}_{i\neq\mu}(0)=\mathbb{I}_{d_{i}}/d_{i}$, where $d_{i}=2j_{i}+1$ is the dimension of the
$i^{\rm th}$ subsystem with total angular momentum quantum number $j_{i}$.
The identity matrices represent the maximally mixed (equal probability) states of the nuclear spin moments of the $^{19}$F$^{-}$ ($I=1/2$) and $^{7}$Li$^{+}$ ($I=3/2$) ions which, together with $\mu^{+}$ ($S=1/2$), gives a total number of states, $N_{s}=\prod_{i=1}^{N_{m}}d_{i}=512$ for each of the two orientations of the $\mu$F$_{2}$Li$_{2}$F$_{2}$ cluster [red and orange stars in Fig.~\ref{fig1}(b)].
The model Hamiltonian is determined by considering a system of $N_{m}$ magnetic moments coupled via long-range magnetic dipole interactions,
\begin{equation}
\hat{H}_{0}=\frac{1}{2}\sum_{i\neq j}^{N_{m}}\frac{\mu_{0}\gamma_{i}\gamma_{j}}{4\pi|{\bm r}_{ij}|^{3}}\left[\hat{\bm J}_{i}\cdot\hat{\bm J}_{j}-3\left(\hat{\bm J}_{i}\cdot{\bm n}_{ij}\right)\left(\hat{\bm J}_{j}\cdot{\bm n}_{ij}\right)\right],
\label{Hsys}
\end{equation}
where ${\bm r}_{ij}={\bm r}_{i}-{\bm r}_{j}$ are the position vectors connecting moments $i$ and $j$, $\mu_{0}$ is the vacuum permeability,
$\gamma_{i}$ are the gyromagnetic ratios \cite{supp},
$\hat{\bm J}_{i}=(\hat{J}_{i,x},\hat{J}_{i,y},\hat{J}_{i,z})$ are the angular momentum vector operators,
${\bm n}_{ij}={\bm r}_{ij}/|{\bm r}_{ij}|$,
and the prefactor ($1/2$) avoids double counting.
Quadrupolar coupling to nuclear spins with $I\geq1$ from finite electric field gradients has recently been discussed in detail \cite{gomilsek:22,bonfa:22},
however, these interactions have been neglected in our model since they are relatively weak for $^{7}$Li ($I=3/2$) so will only affect the $\mu^{+}$
polarization at longer times \cite{wilkinson:20,wilkinson:21}.

For the simpler case of a F--$\mu$--F complex, we need only consider the spin of the implanted $\mu^{+}$ and the two F$^{-}$ nuclear spins.
The energy level spectrum and allowed transitions are shown in Fig.~\ref{fig1}(c) in units of $\hbar\omega_{\mathrm{D}}=\hbar^{2}\mu_{0}\gamma_{\mu}\gamma_{\rm F}/(4\pi \overline{r}_{\mu{\rm F}}^{3})$, where $\overline{r}_{\mu{\rm F}}\approx1.2$~\AA~is the average nearest-neighbor $\mu$--F bond length.
The magnetic field dependence of the F--$\mu$--F eigenstate energies is determined by diagonalizing the Hamiltonian $\hat{H}=\hat{H}_{0}+\hat{H}_{1}$ where $\hat{H}_{1}=-\sum_{i=1}^{N_{m}}\hat{\bm \mu}_{i}\cdot{\bm B}$ is the sum of Zeeman interactions between the $i^{\rm th}$ magnetic moment, $\hat{\bm \mu}_{i}=\gamma_{i}\hat{\bm J}_{i}$, and the applied magnetic field, ${\bm B}$.
For ${\bm B}=0$, the energy eigenstates exist as doubly degenerate pairs because a state with all of its spins reversed has the same energy.
However, one member of each pair is more and the other less likely to be populated because $\mu^{+}$ is initially polarized in the $\left|\uparrow\right\rangle$ state.
The eigenstate vectors of a dipole-coupled, linear ($180^{\circ}$), equidistant F--$\mu$--F complex are listed in Eq.~S3 of Ref.~\cite{supp}.
The on-resonance transitions we are driving can then be written as,
\begin{equation}
\begin{split}
&\left|\uparrow\uparrow\uparrow\right\rangle\rightleftharpoons\bigg(\frac{3+\sqrt{3}}{12}\bigg)^{\frac{1}{2}}\bigg[(1+\sqrt{3})\left|\downarrow\uparrow\uparrow\right\rangle+\left|\uparrow\uparrow\downarrow\right\rangle+\left|\uparrow\downarrow\uparrow\right\rangle\bigg];\\
&\left|\downarrow\downarrow\downarrow\right\rangle\rightleftharpoons\bigg(\frac{3+\sqrt{3}}{12}\bigg)^{\frac{1}{2}}\bigg[(1+\sqrt{3})\left|\uparrow\downarrow\downarrow\right\rangle+\left|\downarrow\downarrow\uparrow\right\rangle+\left|\downarrow\uparrow\downarrow\right\rangle\bigg],
\end{split}
\label{trans}
\end{equation}
where the first spin in each ket label corresponds to $\mu^{+}$ and the spin-quantization axis is parallel to the F--$\mu$--F bond axis.
Note that the eigenstates on the right hand side of Eq.~\ref{trans} are not separable (cannot be written as a tensor product of individual spin states,
$\left|abc\right\rangle\neq\left|a\right\rangle\otimes\left|b\right\rangle\otimes\left|c\right\rangle$)
meaning that the $\mu^{+}$ spin and two F$^{-}$ nuclear spins are entangled and, by resonantly driving transitions of this type, we are properly manipulating entangled state populations.
The energy level structure of the $\mu$F$_{2}$Li$_{2}$F$_{2}$ cluster strongly resembles that of the F--$\mu$--F complex, but appears broadened by the influence of the other moments included in the cluster.
To prevent significant perturbation to the F--$\mu$--F energy levels, the applied excitation always had magnetic field amplitude $0.15\leq B_{y}\leq0.30$~mT [see Fig.~\ref{fig1}(c)].

\begin{figure*}[t!]
\centerline{\includegraphics[width=1.0\linewidth]{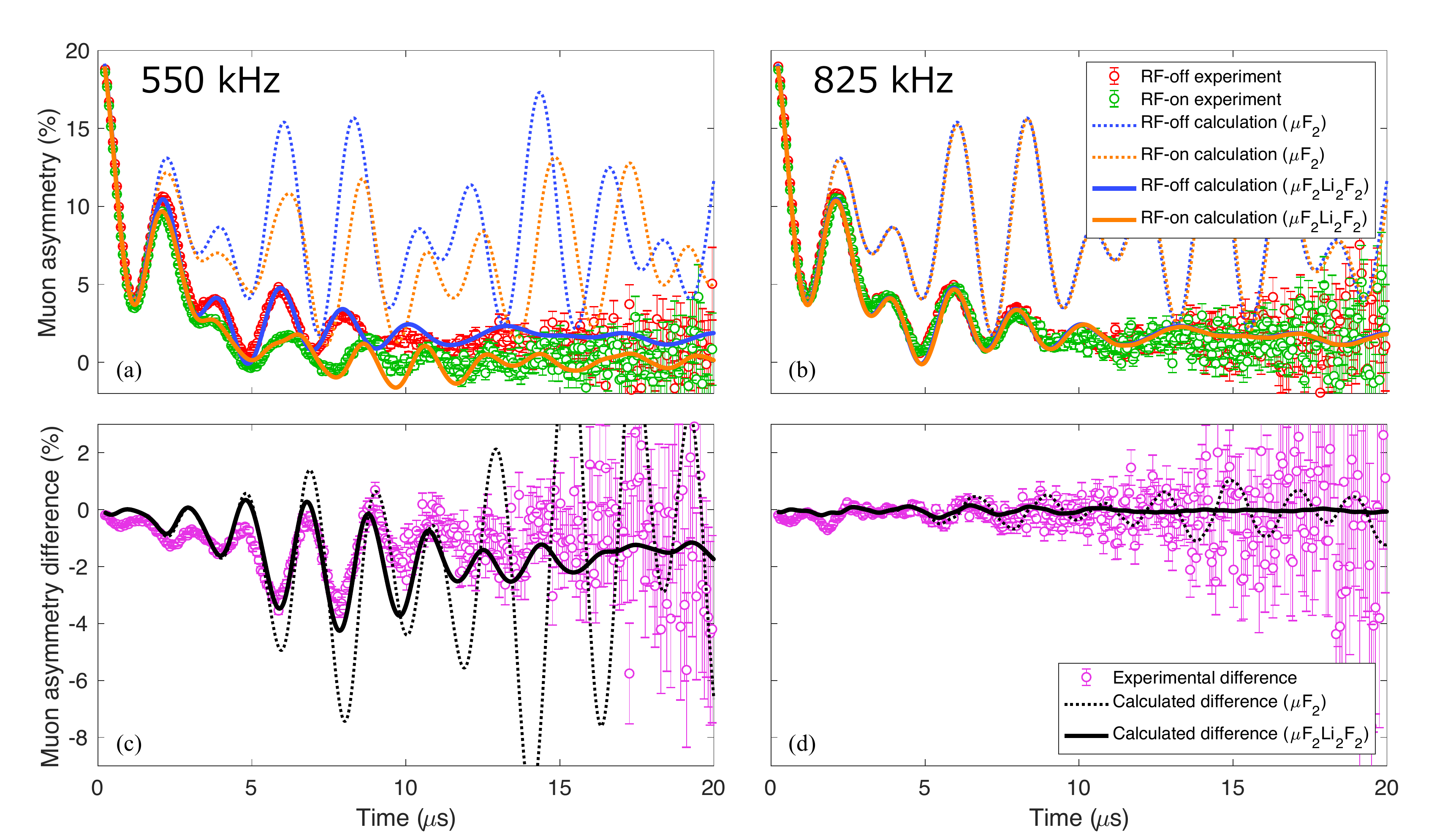}}
\caption{Experimental $\mu^{+}$ asymmetry for LiY$_{0.95}$Ho$_{0.05}$F$_{4}$ at $T=50$~K for both RF-on and RF-off,
at excitation frequencies of (a) $\omega_{\rm on}/(2\pi)=550$~kHz and (b) $\omega_{\rm off}/(2\pi)=825$~kHz,
together with those calculated for the F--$\mu$--F complex ($\mu$F$_{2}$) and the $\mu$F$_{2}$Li$_{2}$F$_{2}$ cluster.
The improved agreement with experiment from accurately modeling the decoherence effect
(by including the local environment of the F--$\mu$--F complex to form the $\mu$F$_{2}$Li$_{2}$F$_{2}$ cluster) is evident.
(c,d) Experimental and calculated differences between the RF-on and RF-off $\mu^{+}$ asymmetries in (a,b), respectively.
Importantly, the on-resonance asymmetry difference cannot be caused by a simple change in relaxation rate.
}
\label{fig2}
\end{figure*}

For a system interacting with a time-dependent magnetic field, ${\bm B}(t)=[0,B_{y}\cos(\omega_{\rm c}t),0]$, we have
$\hat{H}(t)=\hat{H}_{0}+\hat{H}_{1}(t)$ which is the semiclassical, driven Hamiltonian that we employ to evolve (according to the Liouville-von Neumann equation)
the initial density matrix to obtain $\hat{\rho}_{\mu{\rm F_{2}Li_{2}F_{2}}}(t)$ from which we can calculate the $\mu^{+}$ spin polarization for each of the $n=2$ cluster orientations, $P_{\mu,z}^{{\rm calc},n}(t)$, given by,
\begin{equation}
\left(\frac{\hbar}{2}\right)P_{\mu,z}^{{\rm calc},n}(t)=\left\langle\hat{J}_{\mu,z}\right\rangle(t)={\rm Tr}\left[\hat{\rho}^{n}_{\mu{\rm F_{2}Li_{2}F_{2}}}(t)\hat{J}_{\mu,z}\right],
\end{equation}
where the factor of $\hbar/2$ is to convert the expectation value of the $z$-direction $\mu^{+}$ spin projection operator
to a dimensionless polarization.
The calculated polarization to be compared with the experiment is then the average of the two orientations, $P_{\mu,z}^{\rm calc}(t)=[P_{\mu,z}^{{\rm calc},1}(t)+P_{\mu,z}^{{\rm calc},2}(t)]/2$.
The asymmetry function, $A_{\mu,z}^{\rm fit}(t)$, which we fit to the experimental data is,
\begin{equation}
A_{\mu,z}^{\rm fit}(t)=A_{0}P_{\mu,z}^{\rm calc}(t)+A_{1}\exp\left(-\lambda_{1}t\right)+A_{2},
\label{fit}
\end{equation}
where $A_{0}=18.5\%$, $A_{1}=3.4\%$, $\lambda_{1}=0.48$~$\mu$s$^{-1}$ and $A_{2}=-1.2\%$ are the result of simultaneously fitting our model to the $\omega_{\rm on}/(2\pi)=550$~kHz RF-on ($B_{y}\neq0$) and RF-off ($B_{y}=0$) experimental spectra (further details in Ref.~\cite{supp}).
The exponential component only contributes to a fast initial depolarization of the spectra and comes from $\mu^{+}$ not forming F--$\mu$--F within the sample, while the negative value of $A_{2}$ occurs because the inclusion of the RF-coil within the $\mu^{+}$SR sample environment introduces a constant background asymmetry between the forwards and backwards detectors.
Except for the excitation frequency, the same set of parameters were used to calculate the $\mu^{+}$ asymmetry for $\omega_{\rm off}/(2\pi)=825$~kHz.

Fig.~\ref{fig2} shows the experimental and calculated time-evolution of the $\mu^{+}$ asymmetry for both RF-on and RF-off measurements at $\omega_{\rm on}/(2\pi)=550$~kHz and $\omega_{\rm off}/(2\pi)=825$~kHz.
Additionally, the calculated $\mu^{+}$ polarization for an isolated F--$\mu$--F complex with the same tetragonal geometry as the $\mu$F$_{2}$Li$_{2}$F$_{2}$ cluster is shown both with and without the driving term, $\hat{H}_{1}(t)$, demonstrating the decoherence effect of including the {\it local} environment (additional Li$^{+}$ and F$^{-}$ moments in the $\mu$F$_{2}$Li$_{2}$F$_{2}$ cluster) \cite{wilkinson:20,wilkinson:21}.
In terms of the calculations, the combination of muon-site determination, consideration of the $\mu^{+}$ local environment and the semiclassical evolution of an appropriate initial density matrix is essential for describing our experimental data.

Fig.~\ref{fig2}(c) and (d) show the experimental and calculated differences between the RF-on and RF-off measurements at the same RF-field frequencies as (a) and (b), respectively.
There is clearly a significant change in experimental $\mu^{+}$ asymmetry at the on-resonance excitation frequency compared to off-resonance.
Crucially, the observed difference in the spectra cannot be caused by a simple change in relaxation rate, which strongly supports the claim that we are measuring a modification of state populations and not simply the periodic redistribution of the eigenstate energies by the oscillating ${\bm B}(t)$.
Importantly, only the first term in Eq.~\ref{fit} (which is sensitive to the modified eigenstate populations) contributes to the calculated differences.
Since our model Hamiltonian is time-dependent, the transitions are between states separated in energy by $\hbar\omega_{\rm c}$ and, thus,
any structure in the off-resonance [$\omega_{\rm off}/(2\pi)=825$~kHz] calculated difference [Fig.~\ref{fig2}(d)] is simply related to the periodic degeneracy lifting because none of the states in our model are separated by energy $\hbar\omega_{\rm off}$ at any time.
Fig.~\ref{fig2} shows how $\mu^{+}$, coupled via the magnetic dipole interaction to its local environment, is sensitive to the manipulation of the energy eigenstate populations.

Typically, entangled quantum states lose coherence because of dissipative interactions with the environment \cite{zurek:03}.
In our case, there is no dissipation involving Lindblad operators or {\it ad hoc} exponential decay applied to the calculated $\mu^{+}$ spin polarization, meaning we have created a
quantum-entangled spin system that is effectively isolated because the $\mu^{+}$ spin couples very weakly to its {\it external} environment over the lifetime of the measurement
owing to the small energies involved.
When the implanted $\mu^{+}$ stops in the sample, the magnetic dipole interactions entangle the $\mu^{+}$ spin with the other nuclear spins of the cluster, which correlates their behavior.
This allows the $\mu^{+}$ quantum information to be decohered into the {\it local} environment (subsystem of nuclear spins) with the experimentally observable consequence that the $\mu^{+}$ spin polarization is lost after about $10$~$\mu$s as the pure (polarized) $\left|\uparrow\right\rangle$ state evolves into a mixed (unpolarized) state \cite{zurek:03,wilkinson:20}.
Nevertheless, the quantum state of the $\mu$F$_{2}$Li$_{2}$F$_{2}$ cluster maintains coherence over the lifetime of the measurement and, thus, our experiment is properly manipulating entangled state populations by resonantly driving transitions.

In conclusion, we have employed a dual experimental and computational approach to probe, manipulate and model the populations of entangled quantum states in a
system of spins that we excite with continuous RF-fields corresponding to on- and off-resonant excitations.
$\mu^{+}$SR was used to measure the characteristic oscillations in single crystal LiY$_{0.95}$Ho$_{0.05}$F$_{4}$, and changes were observed on application of the RF-fields.
To model these changes, we constructed and optimized a Hamiltonian and demonstrated that it is capable of describing the experimentally observed $\mu^{+}$ spin polarization.
Our methodology (RF coil design, $\mu^{+}$SR experiment, first-principles $\mu^{+}$ stopping site calculation and quantum evolution of the density matrix with QuTiP) opens up the possibility of performing spectroscopy experiments by varying the RF-field frequency and looking for resonances in the difference spectra.
We can also envisage performing pump-probe-style experiments by using the RF-field to change the state population and subsequently measuring the resulting $\mu^{+}$ relaxation.
Additionally, extending the RF-coil design to allow microwave frequencies ($0.3$--$300$~GHz) will permit measurements of electronic (as opposed to spin) phenomena.
This study paves the way for manipulating and understanding a wide variety of complex spin systems with continuous RF-$\mu^{+}$SR.

\begin{acknowledgments}
We gratefully acknowledge the financial support of the UK EPSRC, grant numbers EP/S016465/1 and EP/N024028/1.
The $\mu^{+}$SR experiment was performed with the approval of the Science and Technology Facilities Council (STFC).
We acknowledge the support of the Supercomputing Wales project, which is part-funded by the European Regional Development Fund (ERDF) via Welsh Government, and the facilities of the Hamilton HPC Service of Durham University.
We would like to thank A.~Vaidya (Durham University), A.~Armour (Univeristy of Nottingham), F.~Flicker (Cardiff University), J.S.~Lord (STFC-ISIS) and S.J.~Blundell (University of Oxford) for useful discussions.
Research data from this paper will be made available via doi:xxx.
\end{acknowledgments}


%

\end{document}


\title{Supplemental material: Radio-frequency manipulation of state populations in an entangled fluorine-muon-fluorine system}

\author{David~Billington}
\email{billingtond1@cardiff.ac.uk}
\affiliation{School of Physics and Astronomy, Cardiff University, Queen's Building, The Parade, Cardiff, CF24 3AA, United Kingdom}
\author{Edward~Riordan}
\affiliation{School of Physics and Astronomy, Cardiff University, Queen's Building, The Parade, Cardiff, CF24 3AA, United Kingdom}
\author{Majdi~Salman}
\affiliation{School of Physics and Astronomy, Cardiff University, Queen's Building, The Parade, Cardiff, CF24 3AA, United Kingdom}
\author{Daniel~Margineda}
\affiliation{School of Physics and Astronomy, Cardiff University, Queen's Building, The Parade, Cardiff, CF24 3AA, United Kingdom}
\author{George~J.W.~Gill}
\affiliation{School of Physics and Astronomy, Cardiff University, Queen's Building, The Parade, Cardiff, CF24 3AA, United Kingdom}
\author{Stephen~P.~Cottrell}
\affiliation{ISIS Facility, Rutherford Appleton Laboratory, Harwell Campus, Didcot, Oxon, OX11 0QX, United Kingdom}
\author{Iain~McKenzie}
\affiliation{TRIUMF, Vancouver, V6T 2A3, Canada}
\author{Tom~Lancaster}
\affiliation{Department of Physics, Center for Materials Physics, Durham University, Durham DH1 3LE, United Kingdom}
\author{Michael~J.~Graf}
\affiliation{Department of Physics, Boston College, Chestnut Hill, Massachusetts 02467, USA}
\author{Sean~R.~Giblin}
\email{giblinsr@cardiff.ac.uk}
\affiliation{School of Physics and Astronomy, Cardiff University, Queen's Building, The Parade, Cardiff, CF24 3AA, United Kingdom}

\date{\today}

\maketitle

\beginsupplement

\section{Experimental details}

The $\mu^{+}$SR experiment was performed in an Oxford Instrument continuous flow cryostat.
All measurements presented were taken at temperature $T=50$~K.
The radio-frequency (RF) electromagnetic driving field, ${\bm B}(t)$, was generated by a saddle coil with an entrance aperture greater than the $\mu^{+}$ spot size.
The coil was tuned and matched to a resistance $R=50$~$\Omega$ at the working frequencies of $550$ and $825$~kHz with field strength $0.15\leq B_{y}\leq0.2$~mT for an applied power $P=120$~W.
The experiment was performed in a differential style with five sequential frames for either RF-on or RF-off.
This sequence was repeated until the desired statistics were obtained.
In the former case the RF power was applied continuously for about $16$ $\mu^{+}$ half-lives [$\tau_{\mu}=2.1969811(22)$~$\mu$s] timed to the implantation of each beam pulse (arriving at $50$~Hz) with each pulse containing about $2000$~$\mu^{+}$. 
The experiment can therefore be defined as a continuous RF-pulsed experiment. We chose to measure single crystal LiY$_{0.95}$Ho$_{0.05}$F$_{4}$ as this could be aligned to give a unique direction of the F--$\mu$--F bond, allowing the contribution of neighboring magnetic moments to be calculated.

The experimental $\mu^{+}$ decay asymmetry, $A_{\mu,z}^{\rm exp}(t)$, is given by,
\begin{equation}
A_{\mu,z}^{\rm exp}(t)=\frac{N_{\rm F}(t)-\alpha N_{\rm B}(t)}{N_{\rm F}(t)+\alpha N_{\rm B}(t)},
\end{equation}
where $N_{\rm F}(t)$ and $N_{\rm B}(t)$ are the number of detections in the forward (${\rm F}$) and backward (${\rm B}$) detectors, respectively, and $\alpha$ accounts for systematic differences between the detectors.

\begin{table}[t!]
\caption{Calculated atomic positions for the two crystallographically equivalent orientations of the $\mu^{+}$ stopping site.}
\centering 
\begin{tabular}{c c c c c c} 
\hline\hline 
Orientation & Species, $i$ & $r_{i,{\bm a}}$ (\AA) & $r_{i,{\bm b}}$ (\AA) & $r_{i,{\bm c}}$ (\AA) & $|{\bm r}_{i\mu}|$ (\AA) \\
\hline 
1 & $\mu$ &  2.6417 &  1.3201 &  1.2929 & 0.0000 \\
  & F1    &  3.6870 &  1.5906 &  0.9277 & 1.1398 \\
  & F2    &  1.5773 &  1.0138 &  1.7129 & 1.1846 \\
  & Li1   &  2.9183 & -0.1888 &  3.0538 & 2.3353 \\
  & Li2   &  2.1918 &  2.9164 & -0.4962 & 2.4396 \\
  & F3    &  1.6618 &  1.1420 & -0.9762 & 2.4781 \\
  & F4    &  3.5750 &  1.5075 &  3.6399 & 2.5326 \\
\hline
2 & $\mu$ & -3.9468 &  5.2685 &  4.0078 & 0.0000 \\
  & F1    & -4.2174 &  6.3137 &  3.6426 & 1.1398 \\
  & F2    & -3.6405 &  4.2040 &  4.4277 & 1.1846 \\
  & Li1   & -2.4379 &  5.5450 &  5.7686 & 2.3353 \\
  & Li2   & -5.5431 &  4.8186 &  2.2187 & 2.4396 \\
  & F3    & -3.7687 &  4.2885 &  1.7387 & 2.4781 \\
  & F4    & -4.1343 &  6.2017 &  6.3547 & 2.5326 \\
\hline \hline
\end{tabular}
\label{pos}
\end{table}

\section{Electronic structure calculations}

In order to understand the origin of the contributions to the $\mu^{+}$SR signal, we carried out density functional theory (DFT) calculations in order to locate the most probable $\mu^{+}$ stopping sites and assess the degree of perturbation the $\mu^{+}$ probe causes in the system \cite{moller:13}.
The calculations were carried out using the MuFinder software \cite{mufinder1,mufinder2} and the planewave-based code {\sc Castep} \cite{castep} using the local density approximation.
We used a $2\times2\times1$ supercell in order to minimize the effects of $\mu^{+}$ self-interaction resulting from the periodic boundary conditions.
Muons, modeled by an ultrasoft hydrogen pseudopotential, were initialized in range of low-symmetry positions and the structure was allowed to relax (keeping the unit cell fixed) until the change in energy per ion was less than $2\times10^{-5}$~eV.
An energy cutoff of $490$~eV and a $1\times1\times1$ Monkhorst-Pack grid for ${\bm k}$-point sampling was used.
Table~\ref{pos} lists the calculated atomic positions for the two crystallographically equivalent orientations of the $\mu^{+}$ stopping site.

\begin{table}[t!]
\caption{Fit parameters obtained from simultaneously fitting the $f_{0}=550$~kHz RF-on and RF-off experimental data.}
\centering 
\begin{tabular}{c c} 
\hline\hline 
Parameter                                                                            & $f_{0}=550$~kHz \\
\hline 
$s_{\mu,\rm F12}$~(no units)                                                         &          1.0391 \\
$s_{\mu,\rm Li12}$~(no units)                                                        &          0.8763 \\
$s_{\mu,\rm F34}$~(no units)                                                         &          0.8959 \\
$f_{\rm rel}$~(no units)                                                             &          0.9472 \\
$g_{\rm rel}$~(no units)                                                             &          0.0775 \\
$A_{0}$~(\%)                                                                         &         18.5257 \\
$A_{1}$~(\%)                                                                         &          3.4304 \\
$\lambda_{1}$~($\mu$s)$^{-1}$                                                        &          0.4810 \\
$A_{2}$~(\%)                                                                         &         -1.1786 \\
\hline
$|{\bm r}'_{\mu,{\rm F1}}|=s_{\mu,\rm F12}|{\bm r}_{\rm F1}-{\bm r}_{\mu}|$~(\AA)    &          1.1844 \\
$|{\bm r}'_{\mu,{\rm F2}}|=s_{\mu,\rm F12}|{\bm r}_{\rm F2}-{\bm r}_{\mu}|$~(\AA)    &          1.2309 \\
$|{\bm r}'_{\mu,{\rm Li1}}|=s_{\mu,\rm Li12}|{\bm r}_{\rm Li1}-{\bm r}_{\mu}|$~(\AA) &          2.0464 \\
$|{\bm r}'_{\mu,{\rm Li2}}|=s_{\mu,\rm Li12}|{\bm r}_{\rm Li2}-{\bm r}_{\mu}|$~(\AA) &          2.1378 \\
$|{\bm r}'_{\mu,{\rm F3}}|=s_{\mu,\rm F34}|{\bm r}_{\rm F3}-{\bm r}_{\mu}|$~(\AA)    &          2.2201 \\
$|{\bm r}'_{\mu,{\rm F4}}|=s_{\mu,\rm F34}|{\bm r}_{\rm F4}-{\bm r}_{\mu}|$~(\AA)    &          2.2690 \\
$f_{\rm c}=f_{\rm rel}f_{0}=\omega_{\rm c}/(2\pi)$~(kHz)                             &        520.9730 \\
$E_{\rm c}=\hbar\omega_{\rm c}$~(neV)                                                &          2.1546 \\
$g_{\rm c}=g_{\rm rel}E_{\rm c}$~(neV)                                               &          0.1671 \\
$B_{y}=g_{\rm c}/(\hbar\gamma_{\mu})$~(mT)                                           &          0.2980 \\
\hline
$\chi_{\rm red}^{2}$~(no units)                                                      &         10.1847 \\
\hline \hline
\end{tabular}
\label{fit}
\end{table}

\section{Fitting the asymmetry}

In the fits, we also allow some freedom in the positions of the subsystems relative to the values from first-principles calculations because local-density and gradient-corrected approximations to the exchange-correlation energy functional
are well-known to lead to over- and under-binding, respectively.
This additional freedom is given by a dimensionless parameter, $s_{\mu,i}$, which represents an expansion ($s_{\mu,i}>1$) or contraction ($s_{\mu,i}<1$) of the bond length between the $\mu^{+}$ and the $i^{\rm th}$ species, {\it i.e.} ${\bm r}'_{\mu,i}=s_{\mu,i}{\bm r}_{\mu,i}=s_{\mu,i}({\bm r}_{i}-{\bm r}_{\mu})$.
The driving frequency is also allowed to vary from its nominal value, {\it i.e.} $\omega_{\rm c}=2\pi f_{\rm rel}f_{0}$
where $f_{\rm 0}$ is the nominal driving frequency (the experimental RF-field frequency) and $f_{\rm rel}$ is the
dimensionless fit parameter.
This is valid because the RF coil has a finite $Q$-factor so there will be some component of the magnetic field oscillating at the frequency corresponding to the energy level separation.
The strength of the coupling of the dipole system to the driving field is also allowed to vary, and is calculated
from $g_{\rm c}=g_{\rm rel}E_{\rm c}=g_{\rm rel}\hbar\omega_{\rm c}$, where $g_{\rm rel}$ is the dimensionless fit parameter
that is, essentially, proportional to the magnetic field strength, $B_{y}=g_{c}/(\hbar\gamma_{\mu})$, which controls the transition rate.

For the fits, we employ a differential evolution algorithm \cite{storn:97} to find the global minimum of the reduced chi-squared,
\begin{equation}
\chi_{\rm red}^{2}=\left(\frac{1}{N_{p}-v}\right)\sum_{i=1}^{N_{p}}\bigg(\frac{A_{\mu,z}^{\rm exp}(t_{i})-A_{\mu,z}^{\rm fit}(t_{i})}{\sigma_{A_{\mu,z}^{\rm exp}}(t_{i})}\bigg)^{2},
\end{equation}
where $N_{p}$ is the number of data points and $v$ is the number of fitting parameters.
Beyond $12.5$~$\mu$s, the uncertainty in the experimental $\mu^{+}$ spin polarization becomes large because several $\mu^{+}$ half-lives [$\tau_{\mu}=2.1969811(22)$~$\mu$s] have passed making it difficult to judge the agreement between experiment and calculation and, therefore, $\chi_{\rm red}^{2}$ is only evaluated for $t\leq12.5$~$\mu$s.
Table~\ref{fit} lists the parameters obtained by simultaneously fitting the RF-on and RF-off experimental data at $\omega_{\rm on}/(2\pi)=550$~kHz.
Note that the same set of parameters, but with $\omega_{\rm on}/(2\pi)=550$~kHz replaced with $\omega_{\rm off}/(2\pi)=825$~kHz,
produce the off-resonance $\mu^{+}$ asymmetry in Fig.~2(b,d) of the main text.

\section{Fluorine-muon-fluorine eigenstates}

For a linear ($180^{\circ}$), equidistant F--$\mu$--F complex, diagonalizing the magnetic dipole Hamiltonian \cite{muonbook}
($\hat{H}_{0}$ in the main text, with only the $\mu$--F1 and $\mu$--F2 interactions included) gives four doubly-degenerate
energy eigenstates with eigenenergies [in units of $\hbar\omega_{\rm D}=\hbar^{2}\mu_{0}\gamma_{\mu}\gamma_{\rm F}/(4\pi r_{\mu{\rm F}}^{3})$]
of $E_{1,2}=-1$, $E_{3,4}=(1-\sqrt{3})/2$, $E_{5,6}=0$, $E_{7,8}=(1+\sqrt{3})/2$.
The corresponding eigenvectors (in a $\hat{J}_{z}$-basis, with $z$ along the linear F--$\mu$--F bond axis) are then,
\begin{equation}
\begin{split}
&\left|1\right\rangle=\left|\uparrow\uparrow\uparrow\right\rangle;\\
&\left|2\right\rangle=\left|\downarrow\downarrow\downarrow\right\rangle;\\
&\left|3\right\rangle=\bigg(\frac{3+\sqrt{3}}{12}\bigg)^{\frac{1}{2}}\bigg[(1-\sqrt{3})\left|\downarrow\uparrow\uparrow\right\rangle+\left|\uparrow\uparrow\downarrow\right\rangle+\left|\uparrow\downarrow\uparrow\right\rangle\bigg];\\
&\left|4\right\rangle=\bigg(\frac{3+\sqrt{3}}{12}\bigg)^{\frac{1}{2}}\bigg[(1-\sqrt{3})\left|\uparrow\downarrow\downarrow\right\rangle+\left|\downarrow\downarrow\uparrow\right\rangle+\left|\downarrow\uparrow\downarrow\right\rangle\bigg];\\
&\left|5\right\rangle=\frac{1}{\sqrt{2}}\big(\left|\uparrow\uparrow\downarrow\right\rangle-\left|\uparrow\downarrow\uparrow\right\rangle\big);\\
&\left|6\right\rangle=\frac{1}{\sqrt{2}}\big(\left|\downarrow\downarrow\uparrow\right\rangle-\left|\downarrow\uparrow\downarrow\right\rangle\big);\\
&\left|7\right\rangle=\bigg(\frac{3-\sqrt{3}}{12}\bigg)^{\frac{1}{2}}\bigg[(1+\sqrt{3})\left|\downarrow\uparrow\uparrow\right\rangle+\left|\uparrow\uparrow\downarrow\right\rangle+\left|\uparrow\downarrow\uparrow\right\rangle\bigg];\\
&\left|8\right\rangle=\bigg(\frac{3-\sqrt{3}}{12}\bigg)^{\frac{1}{2}}\bigg[(1+\sqrt{3})\left|\uparrow\downarrow\downarrow\right\rangle+\left|\downarrow\downarrow\uparrow\right\rangle+\left|\downarrow\uparrow\downarrow\right\rangle\bigg],
\end{split}
\end{equation}
where the first spin in each ket label corresponds to $\mu^{+}$.
While the lowest energy (ground) eigenstates $\left|1\right\rangle$ and $\left|2\right\rangle$ are separable
($\left|\uparrow\uparrow\uparrow\right\rangle=\left|\uparrow\right\rangle\otimes\left|\uparrow\right\rangle\otimes\left|\uparrow\right\rangle$)
the other eigenstates are not.
Eigenstates $\left|5\right\rangle$ and $\left|6\right\rangle$ are, in fact, partially separable,
\begin{equation}
\frac{1}{\sqrt{2}}\big(\left|\uparrow\uparrow\downarrow\right\rangle-\left|\uparrow\downarrow\uparrow\right\rangle\big)=\left|\uparrow\right\rangle\otimes\bigg[\frac{1}{\sqrt{2}}\big(\left|\uparrow\downarrow\right\rangle-\left|\downarrow\uparrow\right\rangle\big)\bigg],
\end{equation}
which we can recognize at the tensor product of the $\mu^{+}$ in the $\left|\uparrow\right\rangle$ state and the two fluorines in an entangled (nonseparable)
spin-singlet state (nonmagnetic, $J=0$).
Eigenstates $\left|3\right\rangle$, $\left|4\right\rangle$, $\left|7\right\rangle$ and $\left|8\right\rangle$ are not separable, thus demonstrating entanglement between the three spin-$1/2$ magnetic moments of the F--$\mu$--F complex.
Inclusion of the F1--F2 magnetic dipole interaction only weakly perturbs the superposition of the nonseparable eigenstates ($\left|3\right\rangle$, $\left|4\right\rangle$, $\left|7\right\rangle$ and $\left|8\right\rangle$) but, importantly, they maintain their $a\left|\downarrow\uparrow\uparrow\right\rangle+b\left|\uparrow\uparrow\downarrow\right\rangle+b\left|\uparrow\downarrow\uparrow\right\rangle$ form and are, therefore, robustly nonseparable (entangled).
Inclusion of the local environment of the F--$\mu$--F complex (Li$_{2}$F$_{2}$) has a similar, weakly-perturbing effect because $\hbar\omega_{\rm D}$ remains the dominant energy scale of the $\mu$F$_{2}$Li$_{2}$F$_{2}$ cluster and, by design, this remains true even in the presence on the RF excitation field.


%